\documentclass[manuscript]{acmart} 

\usepackage{hyperref}             % hyperlinks
\usepackage{booktabs}             % professional-quality tables
\usepackage{multirow}             % for multiple rows/cols in tables
\usepackage{amsfonts}             % blackboard math symbols
\usepackage{amsmath}              % maths
\usepackage{amssymb}              % maths symbols
\usepackage{acronym}              % Acronyms
\usepackage[capitalise]{cleveref} % For clever references

\usepackage[utf8]{inputenc}
\usepackage[T2A,LAE,T1]{fontenc}
\usepackage[russian,main=english]{babel}
\usepackage{arabtex}
\usepackage{utf8}

\acrodef{UR}[U2R]{user-to-recipe}
\acrodef{NN}{Neural Network}
\acrodef{ALS}{Alternating Least Squares}
\acrodef{KNN}[\textit{k}-NN]{\textit{k} Nearest Neighbours}
\acrodef{IR}{Information Retrieval}
\acrodef{MAP}[mAP]{Mean Average Precision}
\acrodef{SOTA}[SoTA]{State of the Art}
\acrodef{MI}{Mutual Information}
\acrodef{PMI}{Pointwise Mutual Information}
\acrodef{BO}{Bayesian Optimisation}
\acrodef{NLP}{Natural Language Processing}
\acrodef{LSA}{Latent Semantic Analysis}
\acrodef{CTR}{Click-Through Rate}
\acrodef{CNN}{Convolutional Neural Network}
\acrodef{TFIDF}[TF-IDF]{Term Frequency-Inverse Document Frequency}
\acrodef{SOTA}{State-of-the-art}
\acrodef{CF}{Collaborative Filtering}
\acrodef{CB}{Content Based}
\acrodef{MLR}{Multi-Language Recommendation}    % Acronyms used

\newcommand{\Ib}{\mathbf{I}}

\newcommand{\M}{\mathbf{M}}

\newcommand{\X}{\mathbf{X}}

\newcommand{\Thetab}{\boldsymbol{\Theta}}

\newcommand{\mub}{\boldsymbol{\mu}}

\newcommand{\Eb}{\boldsymbol{\mathrm{E}}}
\newcommand{\Ub}{\boldsymbol{\mathrm{U}}}
\newcommand{\bb}{\boldsymbol{\mathrm{b}}}
\newcommand{\Sb}{\boldsymbol{\mathrm{S}}}

\newcommand{\ie}{{\em i.e.~\/}}
\newcommand{\eg}{{\em e.g.~\/}}

\newcommand{\cf}{{\em c.f.~\/}}

\renewcommand{\lim}{\operatornamewithlimits{lim}}

\newcommand{\arabictext}[1]{\setcode{utf8}\< \small{#1} >}

\copyrightyear{2020}
\acmYear{2020}
\setcopyright{rightsretained} 
\acmConference[RecSys '20]{Fourteenth ACM Conference on Recommender Systems}{September 22--26, 2020}{Virtual Event, Brazil}
\acmBooktitle{Fourteenth ACM Conference on Recommender Systems (RecSys '20), September 22--26, 2020, Virtual Event, Brazil}\acmDOI{10.1145/3383313.3418478}
\acmISBN{978-1-4503-7583-2/20/09}

\begin{document}

\author{Niall Twomey}
\authornote{Corresponding author.}
\affiliation{%
  \institution{Cookpad Ltd}
  \streetaddress{1 Broad Plain}
  \city{Bristol}
  \postcode{BS2 0JP}
  \country{UK}}
\email{niall-twomey@cookpad.com}

\author{Mikhail Fain}
\authornote{Authors listed alphabetically on last name.}
\affiliation{%
  \institution{Cookpad Ltd}
  \streetaddress{1 Broad Plain}
  \city{Bristol}
  \postcode{BS2 0JP}
  \country{UK}}
\email{mikhail-fain@cookpad.com}

\author{Andrey Ponikar}
\affiliation{%
  \institution{Cookpad Ltd}
  \streetaddress{1 Broad Plain}
  \city{Bristol}
  \postcode{BS2 0JP}
  \country{UK}}
\email{andrey-ponikar@cookpad.com}

\author{Nadine Sarraf}
\affiliation{%
  \institution{Cookpad Ltd}
  \streetaddress{1 Broad Plain}
  \city{Bristol}
  \postcode{BS2 0JP}
  \country{UK}}
\email{nadine-sarraf@cookpad.com}

\title{Towards Multi-Language Recipe Personalisation and Recommendation}


\begin{abstract}

Multi-language recipe personalisation and recommendation is an under-explored field of information retrieval in academic and production systems. The existing gaps in our current understanding are numerous, even on fundamental questions such as whether consistent and high-quality recipe recommendation can be delivered across languages. Motivated by this need, we consider the multi-language recipe recommendation setting and present grounding results that will help to establish the potential and absolute value of future work in this area. Our work draws on several billion events from millions of recipes, with published recipes and users incorporating several languages, including Arabic, English, Indonesian, Russian, and Spanish. We represent recipes using a combination of normalised ingredients, standardised skills and image embeddings obtained without human intervention. In modelling, we take a classical approach based on optimising an embedded bi-linear user-item metric space towards the interactions that most strongly elicit cooking intent. For users without interaction histories, a bespoke content-based cold-start model that predicts context and recipe affinity is introduced. We show that our approach to personalisation is stable and scales well to new languages. A robust cross-validation campaign is employed and consistently rejects baseline models and representations, strongly favouring those we propose. Our results are presented in a language-oriented (as opposed to model-oriented) fashion to emphasise the language-based goals of this work. We believe that this is the first large-scale work that evaluates the value and potential of multi-language recipe recommendation and personalisation. 

\end{abstract}

\maketitle

% Suppress page numbers
% \pagenumbering{gobble}

\section{Introduction and Related Work}
\label{section:introduction}

A plurality of complex factors influence the choices we make when deciding on which recipes to cook. The occurrence of allergens in a recipe will render it inappropriate for some users, the personal commitments of other users will lead to total avoidance of some food categories, and for yet others, embarking on short- and long-term dietary campaigns will disturb otherwise steady and predictable eating habits. Indeed, since membership to any categorisation is non-exclusive and transient, we cannot rely on the existence of a uniformly acceptable recipe ranking across any subgroup. In this paper, we concern ourselves with delivering recipe personalisation models that capture individual preferences.

The formative works in recipe recommendation \cite{Freyne2010, Freyne2010b} constructed user-based ingredient preferences from historic recipe interactions and represented recipes by their ingredients. Distances between users and recipes in the ingredient representation space can be used to make recommendations using \ac{KNN}. This is a classic design pattern in the literature \cite{bobadilla2013recommender,lu2015recommender,zhang2019deep}. Because ingredients and recipes are written in un- or semi-structured forms, they are not necessarily amenable to immediate analysis, and recipe normalisation is known to be beneficial for downstream tasks \cite{Salvadora,kitchenette}. Thus, early work employed ontologies or knowledge graphs \cite{paoloijcai,El-Dosuky2012,Dooley2018}, supervised training \cite{Salvadora} and scoring methods \cite{Teng2012} to extract clean elements. Most of these approaches require a labelled set of `canonical' recipe entities (ingredients, tools, skills) and to date have been evaluated on a single language, which presents a problem for scaling the approaches to a multi-language system. 

Extracting the quality and quantity of recipes and ingredients \cite{Teng2012,Silva2019} is a key precursor in many application areas of food computing, including healthy recommendation \cite{TrangTran2018}. The multi-modal aspect of recipes has shown promise in enhancing cooking procedure understanding \cite{yamakata-etal-2020-english} by using auxiliary data such as video \cite{song2016,malmaud-etal-2015-whats} or images \cite{Salvador2019inversecooking,zhang2019categorization}. The existing work on leveraging these modalities for recommendation \cite{Trattner2018, Kawano2013, Trattner2019a} uses established pre-trained image models or specific image features (including measures of sharpness, contrast). Not all of the generic image representation approaches are suitable for recipe image datasets, which typically consist of images along with semi-structured text (title, ingredients, steps). Thus, transfer learning \cite{Chen2018crossmodal}, cross-modal training \cite{Carvalho2018,wang2019learning,Fu_2020_CVPR}, as well as self-supervised training with weak labels \cite{fain2019dividing}, have been proposed for building recipe image representation.

The application of recipe recommendation models to more than one language is not unexplored \cite{Maia2018, Trattner2018}, though the scope of these approaches is limited to two languages and is reliant on manual intervention for recipe pre-processing. Consequently, the limits of large-scale multi-lingual recipe modelling are under-explored, despite the existence of several relevant datasets and platforms operating in several languages \cite{ harashima-etal-2016-large, Majumder2020}. \ac{SOTA} neural translation \cite{wu2016google, bahdanau2014neural} and multi-language \ac{NLP} frameworks \cite{qi2020stanza} offer opportunities for bridging these gaps. To the best of our knowledge, ours is the first comprehensive work delivering intrinsic language-agnostic pipelines for recipe recommendation and personalisation across many languages. Towards this end, our methodological contributions and results are outlined in \cref{section:methods,section:results}, and we conclude on the value, limitations and future directions of this research in \cref{section:conclusions}.

\section{Methods} \label{section:methods}

\subsection{Dataset}
\label{section:interactions}

We consider only the published recipes with valid titles, ingredient lists, ingredient quantities, and method steps (as well as optional fields cooking duration, serving size, images). We limit ourselves to a single online multi-language recipe platform (Cookpad) as we are unaware of alternative data sources fulfilling our multi-language requirements. 

We employ ten different user interaction types, and each is assigned a weight based on its likelihood of indicating future cooking behaviour estimated from proprietary data. 
Search data are also considered. Cookpad's search seeks to serve the best \textit{new} recipes to users based on their query. We have access to the queries, search result order and recipe clicks. Data fusion techniques are incorporated to merge the interaction data to the click data arising from the search histories in our analyses.

\subsection{Data Representations}\label{section:recipe_embeddings}

\subsubsection{Qualitative Features}
The purpose of these deterministic features is to provide key insights into recipe complexity, completeness, quality, and regionality. Specifically, we extract the following features: 1. the number of ingredients used; 2. the number of skills used; 3. recipe image; 4. the number of steps; 5. the number of step images; 6. the ratio of steps to step images; 7. the published year; 8. the published month; 9. the published time; 10. the author's identity; 11. the author's system ID 12. the cooking time; 13. the number of portions; 14. author's country.

\subsubsection{Normalised Ingredients}

Recipe data contains ingredients with quantities in separate fields in various languages. Quantity extraction quality is not uniform across datasets with certain regional traits, while the ingredients themselves are written in raw form. For each language, we split the dataset by spaces, and, ignoring numerical values, extract a set of 200 common quantity tokens. We then remove any ingredients with punctuation or quantity present, and sort the remaining ingredients by frequency, picking the top 1500 common ingredients as the dictionary. 

For inference, the raw ingredient string is matched to an ingredient from the dictionary by tokenising it and finding the largest common subset of tokens between a candidate normalised ingredient and the original ingredient. To deal with misspellings we use a threshold on the cosine distance between word vectors trained on a recipe corpus using FastText \cite{bojanowski2016enriching}. Our algorithm is unsupervised and can be applied to all languages with space separation between words.

\subsubsection{Normalised Skills}\label{section:skill_discovery}
Pre-trained \ac{SOTA} language models \cite{qi2020stanza} are deployed to detect verbs in recipe sentences. For a given sentence, the cartesian product between the sets of detected verbs and ingredients defines all possible ingredient-verb pairs for that sentence, \eg `slice' and `onion'. We incorporated partial matching and Levenshtein distance to overcome ingredient naming inconsistencies (\eg when `peppers' in a recipe step refers to `red bell peppers' from the ingredient list). We broadcast ingredient-verb pair extraction over our datasets, extending existing work in skill extraction \cite{zhang2019categorization}.

Let $C_I(i)$, $C_V(v)$, $C_{I,V}(i, v)$, $C_{I|V}(i|v)$ and  $C_{V|I}(v|i)$ denote a family of counting functions for ingredients, verbs, joint ingredient-verb pairs, and conditional ingredient-verb pairs in recipe sentences. The domain of the random variables are $I \in \{i, \lnot i\}$ and $V \in \{v, \lnot v\}$ (\ie `present' and `absent'). Removing subscripts for improved clarity, counts are normalised to form distributions (\eg $P(i, v) = C(i,v) \big/ \sum_{i', v'} C(i', v')$), allowing us to calculate \ac{MI} ($\sum_{i, v} P(i, v) \log \frac{ P(i, v) }{ P(i) P(v) } $). The summands are reformulated in terms of positive ingredient and verb occurrences since only these counts are available, \ie $P(\lnot a) = 1 - P(a)$ and $P(a, \lnot b) = \left(1 - P(b|a)\right) P(a)$. Denoting the \ac{MI} matrix as $\M \in \mathcal{R}^{|I| \times |V|}$, the expected \ac{MI} of ingredient $i$ over associated verbs as $\mathbb{E}_{v' \sim P_{V|I=i}}\left[ \M_{i, v'} \right]$, and the expected \ac{MI} of verb $v$ over ingredients as $\mathbb{E}_{i' \sim P_{I|V=v}}\left[ \M_{i', v} \right]$, the data-dependent threshold for the $(i, v)$-th ingredient-verb pair is defined as

\begin{align}
    \Thetab_{i, v} = \alpha \mathbb{E}_{v' \sim P_{V|I=i}}\left[ \M_{i, v'} \right] + (1 - \alpha ) \mathbb{E}_{i' \sim P_{I|V=v}}\left[ \M_{i', v} \right] \nonumber
\end{align}

\noindent where $\alpha \in [0, 1]$ is a hyperparameter (default value 0.5) that balances the relative weight of ingredients and verbs in skill selection. The final set of ingredient-skill pairs is given by $V = \{(i, v) : \M_{i,v} > \Thetab_{i, v} ~ \forall ~ i, v\}$.

\subsubsection{Text Representations}\label{section:text_representation}
We explore representing recipe text as a bag of sub-word-units with \ac{TFIDF} embeddings. The dimensionality of the embedding was reduced to 300 using singular value decomposition, and we followed the FastText \cite{bojanowski2016enriching} procedure in sub-word unit selection.

\subsubsection{Image Representations}\label{section:image_representation}
We used self-supervised training to extract image representations \cite{fain2019dividing}, extending the method to the multilingual setting. We trained a DenseNet-201 \cite{Huang} model using a variation of TagSpace \cite{Weston}. According to this approach, TagSpace labels in all languages and images are all embedded in the same shared 300-dimensional space, which leads to similar labels (in the same or different languages) ending up close to each other in the shared embedding space.   
The labels for training were the top 1000 common unigrams and bigrams extracted from recipe titles per each of the 5 languages. After the model was trained for 40 epochs on a dataset with 1.5M images and 5k labels, we discard the label embeddings and run the CNN on all recipe images.  The image representations are extracted from the global average pooling layer after the last convolutional layer.

\subsubsection{User Representations}\label{section:user_profile}
In this work, user profiles derive directly from users' interaction histories. Let $\Ib \in \mathbb{R}_+^{N_u \times N_r}$ be the (sparse) user-to-recipe interaction matrix that encodes the interaction importance numerically. Moreover, let $\mub$ be the row-wise normalised interaction matrix, \ie $\mub_{u, r} = \Ib_{u, r} / \sum_{r'} \Ib_{u, r'} ~~ \forall u, r$. Finally, let the recipe features be embedded in $\X \in \mathbb{R}^{N_r \times D}$ (\cf \cref{section:recipe_embeddings}). Using these definitions, the user features are calculated simply by averaging recipe embeddings over interaction history, \ie $\Ub = \mub \X$, and $\Ub \in \mathbb{R}^{N_u \times D}$. 

\subsection{Behavioural Models}

\subsubsection{`Clickability' Model}
We used a 3-layer feed-forward neural network with ReLU non-linear activation units, dropout, and batch normalisation as a cold-start. 
The model is optimised in a pairwise approach \cite{pasumarthi2019tf} to differentiate between positive (clicked recipes) and negative (recipes not clicked) recipes. Negatives are sampled randomly from among the recipes viewed in the search results but not clicked. We incorporate late context fusion to measure the impact of using query context in ranking. We call this the `clickability' model since it is based on click-through data.

\subsubsection{Personalisation Model}

LightFM \cite{Kula2015a} is a framework offering linear \ac{CF}, \ac{CB} and hybrid recommendation models. It is known to be strongly performant in scenarios with sparse and transient data, even for new users with little interaction history \cite{Kula2015a}. Our objective in this research is to establish strong definitive baselines for multi-language recipe recommendation \cite{dacrema2019we}. Consequently, our future work will develop the assessment of \ac{SOTA} recommendation models in this application area. 
In LightFM's setting, given user and recipe embedding matrices ($\Eb^U \in \mathbb{R}^{D_U \times K}$ and $\Eb^R \in \mathbb{R}^{D_R \times K}$), user and recipe embeddings are calculated with $\X^U=\Ub \Eb^U$ and $\X^R=\X \Eb^R$. User-recipe affinity is measured by $\Sb_{u, r} = f\left( \X^U_u \cdot \X^R_r + \bb^U_u + \bb^R_r \right)$ where $\bb^U$ and $\bb^R$ are user- and recipe-specific biases, and $f(\cdot)$ is a suitable function selected based on the task, \eg logistic. In order to optimise the embedding matrices and biases, several hyperparameters must be specified (including learning rate, number of iterations, user and recipe regularisation, the loss function, sample weights, feature groups). Owing to the large number of hyperparameters, we take a sequential approach and use \ac{BO} \cite{mockus2012bayesian} on a validation set to select these parameters. 

\subsection{Evaluation Procedures}\label{section:experiments}

\begin{table}[h]
    \centering
    \caption{The approximate number of users, recipes and interactions available for analysis. 
    }
    \begin{tabular}{lcccccc}
        \toprule
                      & Arabic & English & Indonesian & Russian & Spanish & Total \\
         \midrule
         Users        &   2M &   3M & 6M &   2M &   4M & 18M \\
         Items        & 0.6M & 0.5M & 2M & 0.5M & 0.5M & 4M  \\
         Events       & 0.8B & 0.2B & 4B & 0.5B &   1M & 7B  \\
         \bottomrule
    \end{tabular}
    \label{table:dataset_description}
\end{table}

\Cref{table:dataset_description} presents dataset size that is available to us in this work. We stratify interaction data based on event time into four non-overlapping partially ordered sets named profile, train, validation and test, denoted $S_p \prec S_t \prec S_v \prec S_e$. 
Of particular importance is the `profile' interactions ($S_p$) since these are exclusively used in creating user profiles (\cf \cref{section:user_profile}). 
Performance is evaluated with \ac{MAP} \cite{su2015relationship} at $k \in \{1, 20\}$. Since in this work we consider a lot of moving parts for our system, we opted to rely on \ac{BO} in assessment rather than on ablation studies. 

\section{Results and Discussion}
\label{section:results}

\subsection{Ingredient and Skill Validation}

Our ingredient normalisation algorithm picks up non-trivial patterns in the data. For example, the phrase `large sweet strawberries' is normalised to `strawberry', and `large sweet potatoes' gets normalised to `sweet potato'. It is noteworthy that these two similar phrases are both correctly normalised in different ways by our unsupervised model. 
From a quantitative point of view, ingredient normalisation can match and outperform systems based on manually maintained ingredients dictionaries across the five languages considered. We compare performance of our ingredient normaliser to a system built on top of a professionally-maintained proprietory ingredient dictionary provided by Cookpad in \cref{table:normaliser_evaluation}. This shows statistically significant error rates on the dedicated test sets for our normaliser. 

\begin{table}[h]
\caption{Normalisation error rates evaluated by native speakers on each language. Statistical significance results in bold.}
\label{table:normaliser_evaluation}
\begin{center}
\begin{tabular}{l c c c c c}  
\toprule
Language & Arabic & English & Indonesian & Russian & Spanish \\ 
\midrule
Baseline & 0.4 & 0.18 & 0.18 & 0.14 & 0.21 \\ 
Proposed & \textbf{0.26} & \textbf{0.10} & 0.12 & \textbf{0.05} & \textbf{0.09} \\ 
\midrule
Reduction & 0.35 & 0.44 & 0.33 & 0.66 & 0.57 \\
\bottomrule
\end{tabular}
\end{center}
\end{table}

\Cref{table:skill_examples} shows discovered skills from all languages, omitting general skills such as `add' and `mix'. Focusing specifically on English, the proposed pairings are of high quality and diversity (\ie `deboning fish' is not a skill that all cooks will employ). The skill quality on the remaining languages are similar in nature to English, and translations of `peel potato' and `boil water' can be found in the non-English columns of the table. 
\begin{table*}[h]
    \centering
    \caption{An example of the discovered skill-ingredient pairs across five different languages.}
    \begin{tabular}{cccccccccc}
         \toprule
         \multicolumn{2}{c}{Arabic} & \multicolumn{2}{c}{English} & \multicolumn{2}{c}{Indonesian} & \multicolumn{2}{c}{Russian} & \multicolumn{2}{c}{Spanish} \\
         \cmidrule(lr){1-2} \cmidrule(lr){3-4} \cmidrule(lr){5-6} \cmidrule(lr){7-8} \cmidrule(lr){9-10}
         Ing. & Skill & Ing. & Skill & Ing. & Skill & Ing. & Skill & Ing. & Skill  \\
         \midrule
       \arabictext{بيض} & \arabictext{خفق} & onion     & slice  & air  & didihkan & \foreignlanguage{russian}{лук}          & \foreignlanguage{russian}{нарезать} & cebolla & cortar  \\
       \arabictext{دقيق} & \arabictext{خلط} & fish     & debone & telur   &aduk  & \foreignlanguage{russian}{морковь}      & \foreignlanguage{russian}{натереть} & agua    & hervir  \\
       \arabictext{ثوم} & \arabictext{قطع} & eggs      & beat   & margarin  &  panaskan & \foreignlanguage{russian}{разрыхлитель} & \foreignlanguage{russian}{просеять} & cebolla & pelar   \\
       \arabictext{ماء} & \arabictext{غلي} & potatoes & peel   & gula pasir  &  aduk & \foreignlanguage{russian}{картофель}    & \foreignlanguage{russian}{очистить} & ajo     & picar   \\
       \arabictext{دجاج} & \arabictext{قطع} & flour    & sieve  & tepung terigu  & sajikan & \foreignlanguage{russian}{сыр}          & \foreignlanguage{russian}{посыпать} & harina  & amasar  \\
         \bottomrule
    \end{tabular}
    \label{table:skill_examples}
\end{table*}
Since skills are unstructured and may be ambiguous or unclear, we designed a small labelled experiment to evaluate definitive skill detection performance. Training data were acquired using a set of regular expressions and human validation on recipe title, ingredients and steps, and logistic regression models were optimised to detect the chosen skills using a bag of words encoding for each recipe field as features. The dataset contains $\approx30$ times more negatives than positives, yet our method has precision of approximately 0.5. This indicates that skills can be detected with reasonable precision with straightforward approaches.

\subsection{Case Study 1: Interactions}
Since the clickability model is aimed at users without interaction histories, it is trained with click-through data. Additionally, the text and image embeddings from \cref{section:text_representation} were used for recipe representations, and a grid search was used to select the model's hyperparameters (learning rate, network architecture, dropout). For personalisation models, \ac{BO} was used to select model hyperparameters but also to specify the model type (from \ac{CF}, \ac{CB}, or hybrid models) and recipe representation (from any combination of qualitative, ingredient or skill features). We ran \ac{BO} for 100 iterations and selected the optimal model based on the performance on validation sets. 

Results for both models are presented in \cref{table:results_interaction}. High consistency of selected models and representations are obtained, providing evidence that supports ingredients and skills in recommendation. Additionally, hybrid models are always selected over \ac{CF} and \ac{CB} by \ac{BO}. The performance gap between clickability and personalisation is due to two main factors. Firstly, since we sample negatives from the recipes that received new interactions during the test period, it is likely that most of these recipes are of high quality (\ie `clickable') making the task more challenging for content-based models. Secondly, the pool of negatives covers a diverse set of cuisines, which makes recommendation easier for models that have learnt users preferences. Hybrid models are always selected by \ac{BO} which consistently rejected \ac{CF} and \ac{CB} alternatives, suggesting that both preference and content are important for the task.

\begin{table*}[t]
    \centering
    \caption{Results of interaction prediction. \ac{CF} and \ac{CB} baseline results are not shown since they were rejected by \ac{BO}.}
    \begin{tabular}{lcccccccccc}
        \toprule
        \multirow{2}{*}{Language} & \multicolumn{3}{c}{Model spec.}  & \multicolumn{2}{c}{Random model} & \multicolumn{2}{c}{Clickability} & \multicolumn{2}{c}{Personalisation} \\
                  \cmidrule(lr){2-4} \cmidrule(lr){5-6} \cmidrule(lr){7-8} \cmidrule(lr){9-10}
                   & Model  & Ings.      & Skills     & mAP@1 & mAP@20 & mAP@1 & mAP@20 & mAP@1 & mAP20 \\
        \midrule
        Arabic     & Hybrid & \checkmark & \checkmark & 0.103 & 0.246  & 0.126 & 0.272  & 0.397 & 0.459 \\ 
        English    & Hybrid & \checkmark & \checkmark & 0.104 & 0.248  & 0.159 & 0.309  & 0.280 & 0.373 \\ 
        Indonesian & Hybrid & \checkmark & \checkmark & 0.090 & 0.236  & 0.123 & 0.272  & 0.407 & 0.485 \\ 
        Russian    & Hybrid & \checkmark & \checkmark & 0.113 & 0.258  & 0.120 & 0.280  & 0.369 & 0.451 \\ 
        Spanish    & Hybrid & \checkmark & \checkmark & 0.099 & 0.245  & 0.140 & 0.286  & 0.408 & 0.474 \\ 
        \bottomrule
    \end{tabular}
    \label{table:results_interaction}
\end{table*}

We experimented with a variety of alternate text embeddings to understand the source of our good personalisation performance. We found that normalised ingredient and skill were still constantly selected by \ac{BO} even when other text embeddings were available for consideration. This indicates that for the recommendation task defined, targeted ingredient and skill representations are more expressive than general text embeddings.

\subsection{Case Study 2: Search}
In this case study, models are tasked to re-rank candidate recipe lists that were served from search queries. The served search order is known to be significantly biased \cite{joachims2017accurately, wang2018position}, and consequently we expect them to act as an upper bound on performance. Four rankings are considered in this case study: served order (de-biased), clickability, personalisation and the biased served order (biased). Since recipe publication is a random process and served search order is currently a strong function of recency, we can de-bias served results with randomisation. This establishes an unbiased baseline for evaluating clickability and personalisation models. \ac{BO} again selects between \ac{CF}, \ac{CB} and hybrid models.

\Cref{table:results_search} presents the results of the search re-ranking experiment. We base our clickability results on the context-free model variation. We found that adding the query context to ranking models does not substantially improve performance on these metrics since the candidates presented to the model necessarily encapsulate this context already. Our clickability model out-performs baseline significantly and improved performance is obtained over all languages. 

Personalisation out-performs clickability models in all cases, with average \ac{MAP}@1 improvements of $\approx 20\%$.  % (in the range of $\approx 5-30\%$). 
This is a vital metric in search and measures the proportion of time users engage with the top-ranked recipe. English is the weakest language for search personalisation, though it still out-performs baseline and clickability, and Arabic registers the highest improvement over clickability. The interaction experiment has higher base results than search. Although several factors contribute to this, the key explanation is that re-ranking small sets of (potentially) similar candidates for search is more challenging because candidate diversity is lower. 

\begin{table*}[t]
    \centering
    \caption{Results on search re-ranking. \ac{CF} and \ac{CB} baseline results are not shown since they were rejected by \ac{BO}.}
    \begin{tabular}{lccccccccccccc}
        \toprule
        \multirow{2}{*}{Language} & \multicolumn{2}{c}{Served (de-biased)} & \multicolumn{2}{c}{Clickability} & \multicolumn{2}{c}{Personalisation} & \multicolumn{2}{c}{Served (biased)} \\
                  \cmidrule(lr){2-3} \cmidrule(lr){4-5} \cmidrule(lr){6-7} \cmidrule(lr){8-9}
                    & mAP@1 & mAP@20 & mAP@1 & mAP@20 & mAP@1 & mAP20 & mAP@1 & mAP20 \\
        \midrule
        Arabic      & 0.096 & 0.220  & 0.169 & 0.291 & 0.220 & 0.340  & 0.332 & 0.415 \\  % 
        English     & 0.100 & 0.218  & 0.185 & 0.314 & 0.205 & 0.354  & 0.273 & 0.404 \\  % 
        Indonesian  & 0.112 & 0.234  & 0.186 & 0.315 & 0.212 & 0.334  & 0.289 & 0.413 \\  % 
        Russian     & 0.109 & 0.242  & 0.180 & 0.305 & 0.208 & 0.341  & 0.281 & 0.382 \\  % 
        Spanish     & 0.108 & 0.230  & 0.185 & 0.313 & 0.217 & 0.339  & 0.287 & 0.400 \\  % 
        \bottomrule
    \end{tabular}
    \label{table:results_search}
\end{table*}

When evaluating search re-ranking against recipe clicks, we were unable to surpass the strong bias of served order. However, if instead we evaluate performance against other interactions (\eg bookmark, cookplan) the personalisation models out-perform the (biased) served order by $\approx 10\%$. Personalisation models surpassing the strong bias of served search order is noteworthy and highlights the appropriateness of our approach to re-rank search results meaningfully.

\subsection{Case Study Summary and Discussion}

We tested our models extensively against several popular and competitive baseline methods (including \ac{CF} and \ac{CB}) and our proposed approach was exclusively selected by \ac{BO} in interaction and search case studies. Popular text embedding models were also tested, but, disappointingly, these did not increase performance due to averaging effects over long recipe text. This exemplifies the value of targeted recipe representations in recipe recommendation. We focused on reporting qualitative performance measures in this emerging work, and broader measures (including coverage, qualitative) will be factored into more mature future presentations. The prime enabler of our success is the deliberate integration of \ac{SOTA} language models and targeted ingredient and skill recipe representations.

\section{Conclusions}\label{section:conclusions}

The express objective of this paper was to develop initial understanding and expectations in multi-language recipe recommendation. Our analysis, using the most extensive dataset available for cooking and recipe recommendation, validates all representations and models with our results suggesting that multi-language recipe recommendation is suitably modelled with the proposed methodology. Despite this early work employing linear models for personalisation, our approach significantly outperforms popular content-based and collaborative baselines. We believe that we have established a strong standard for comparing the absolute value of succeeding multi-language recommendation research, and our future work will expand into three key areas. First, we will deploy our models to production systems and measure the utility of our methods in live experiments. We will then embark on an exploration of sophisticated non-linear neural recommendation frameworks and evaluate their merit. Finally, we will explore end-to-end cross-language recipe recommenders.

\bibliographystyle{ACM-Reference-Format}
\bibliography{main}

\end{document}